\documentclass[twocolumn,10pt]{article} %
\usepackage[square,numbers,sort&compress,comma]{natbib}

\usepackage{amsmath,bm}
\usepackage{amssymb}
\usepackage{caption}
\usepackage{graphicx}
\usepackage{latexsym}
\usepackage{times}
\usepackage{subcaption}
\usepackage{soul}
\usepackage{xcolor}
\newcommand{\change}[1]{\textcolor{black}{#1}}

\topmargin - 12pt %
\renewenvironment{abstract}%
 {%
  \small%
  {\bfseries \abstractname}%
  \par%
  \vspace{10pt}%
  }%

\renewcommand\abstractname{Abstract}
\newcommand{\nomenclature}%
  [1]%
  {%
  \bgroup%
  \flushleft%
  \small\bf%
  #1%
  \par%
  \egroup%
  }%

\renewcommand{\section}%
  [1]%
  {%
  \bgroup%
  \flushleft%
  \small\bf%
  \stepcounter{section}%
  \arabic{section}. #1%
  \par%
  \egroup%
  }%

\renewcommand{\subsection}%
  [1]%
  {%
  \bgroup%
  \flushleft%
  \small\em%
  \stepcounter{subsection}%
  \arabic{section}.%
  \arabic{subsection}. #1%
  \par%
  \egroup%
  }%

\renewcommand{\subsubsection}%
  [1]%
  {%
  \bgroup%
  \flushleft%
  \small\em%
  \stepcounter{subsubsection}%
  \arabic{section}.%
  \arabic{subsection}.%
  \arabic{subsubsection}. #1%
  \par%
  \egroup%
  }%

 \newcommand{\acknowledgement}%
  [1]%
  {%
  \bgroup%
  \flushleft%
  \small\bf%
  #1%
  \par%
  \egroup%
  }%

 \newcommand{\sectionbib}%
  [1]%
  {%
  \bgroup%
  \flushleft%
  \small\bf%
  #1%
  \par%
  \egroup%
  }%

\begin{document}

\title{\LARGE Low Mach number lattice Boltzmann model for turbulent combustion: flow in confined geometries}

\author{{\large Seyed Ali Hosseini$^{a,b,*}$, Nasser Darabiha$^{c}$, Dominique Th\'evenin$^{a}$}\\[10pt]
 {\footnotesize \em $^a$Laboratory of Fluid Dynamics and Technical Flows, University of Magdeburg ``Otto von Guericke'', D-39106 Magdeburg, Germany}\\[-5pt]
 {\footnotesize \em $^b$Department of Mechanical and Process Engineering, ETH Z\"urich, 8092 Z\"urich, Switzerland}\\[-5pt]
 {\footnotesize \em $^c$Laboratoire EM2C, CNRS, CentraleSup\'elec, Universit\'e Paris-Saclay, 3 rue Joliot Curie, 91192, Gif-sur-Yvette Cedex, France}}
\date{}

\small
\baselineskip 10pt

\twocolumn[\begin{@twocolumnfalse}
\vspace{50pt}
\maketitle
\vspace{40pt}
\rule{\textwidth}{0.5pt}
\begin{abstract} %
 A hybrid lattice Boltzmann/finite-difference solver \change{for low Mach thermo-compressible flows} developed in earlier works is extended \change{to more realistic and challenging configurations involving turbulence and complex geometries} in the present article\change{. The major novelty here as compared to previous contributions is the application of a more robust collision operator, considerably extending the stability of the original single relaxation time model and facilitating larger Reynolds number flow simulations. Additionally,} a subgrid model and the thickened flame approach \change{have also been added allowing for} efficient \change{large eddy} simulations of turbulent reactive flows in complex geometries. This robust solver, in combination with appropriate treatment of boundary conditions, is used to simulate combustion in two configurations: flame front propagation in a 2-D combustion chamber with several obstacles, and the 3-D PRECCINSTA swirl burner. Time evolution of the flame surface in the 2-D configuration shows very good agreement compared to direct numerical and large eddy simulation results available in the literature. The simulation of the PRECCINSTA burner is first performed in the case of cold flow using two different grid resolutions. Comparisons with experimental data reveal very good agreement even at lower resolution. The model is then used, with a 2-step chemistry and multi-component transport/thermodynamics, to simulate the combustor at operating conditions similar to previously reported experimental/numerical studies for \change{$\phi$=0.83}. Results are again in very good agreement compared with available large eddy simulation results as well as experimental data, demonstrating the excellent performance of the hybrid solver.
\end{abstract}
\vspace{10pt}
\parbox{1.0\textwidth}{\footnotesize {\em Keywords:} Combustion; Turbulence; Swirl burner; Large eddy simulation; Lattice Boltzmann method}
\rule{\textwidth}{0.5pt}
\vspace{10pt}
\end{@twocolumnfalse}] 

\clearpage
\section{Introduction}
The lattice Boltzmann method (LBM) is a numerical tool originally intended and now widely used for the incompressible flow regime. It provides an interesting alternative to classical solvers as it does not involve non-linear and non-local discrete operators and relies on a strictly local closure for the pressure field making it highly adapted to parallel processing. The approach has been extended to a plethora of more complex physics.

However, despite a few publications~\cite{yamamoto_simulation_2002,filippova_novel_2000}, development of LBM solvers for combustion and multi-species reacting flows was very slow until recently~\cite{hosseini_hybrid_2019,tayyab_hybrid_2020,sawant_consistent_2021}. A number of different approaches have been proposed over the past couple of years which, along with other publications from the early 2000's, can be categorized as: (a) Hybrid, or (b) pure LBM. The first category makes use of the LBM only for the momentum/pressure/density fields and relies on classical solvers for the energy and species fields, while the pure LBM uses a lattice Boltzmann solver per variable leading to more pronounced memory/computation overhead. Furthermore, different closures for the pressure field can be used: low Mach formulation relying on modified pressure fluctuation/momentum lattice Boltzmann solver~\cite{filippova_novel_2000,hosseini_hybrid_2019} and fully compressible formulation~\cite{tayyab_hybrid_2020,sawant_consistent_2021} which can be subject to entropy/vorticity mode coupling leading to spurious currents around large temperature gradients~\cite{renard_linear_2021}.

While publications targeting combustion using LBM have risen over the past few years, simulation of realistic configurations with complex geometries and turbulent flows remain limited~\cite{hosseini_low-mach_2020,tayyab_lattice-boltzmann_2021}. Robustness of the numerical scheme along with proper treatment of complex geometries are two ingredients needed to perform such simulations. 

In the present work, a previously-introduced model for low Mach combustion~\cite{hosseini_hybrid_2019,hosseini_development_2020} is extended via a more robust realization using a multiple-relaxation frame-work \change{relying on Cumulants of the distribution function. The realization presented here is different from the original form of the Cumulants collision operator in the sense that it is applied to a pressure/momentum LBM. In addition,} a subgrid scale model accounting for under-resolved flow structures, a flame thickening module and proper treatment of curved boundaries for both lattice Boltzmann and finite-difference solvers \change{are used}. The resulting scheme is then used to perform simulations in two complex geometries: deflagration in a 2-D chamber with obstacles, and cold and reactive flows in a realistic swirl burner.

Section \ref{sec:Equations} 2 \change{briefly reviews} the lattice Boltzmann method used to recover the target macroscopic equations and \change{introduces} the multiple-relaxation realization.
In Sec. \ref{sec:test_case} 3, the numerical scheme is used to first model deflagration in a 2-D chamber with obstacles and results are compared to data from both direct numerical (DNS) and large eddy simulations (LES). The section ends with simulations of the PRECCINSTA swirl burner, both for cold and reactive flows. The obtained results are compared to numerical results from LES as well as experimental data from Laser Doppler Anemometry (LDA) measurements. 

\section{\label{sec:Equations} Brief overview of models and implementation}
\subsection{Lattice Boltzmann model}
To solve the \change{low Mach} aerodynamic equations, we use a lattice Boltzmann model that we have developed in previous works \cite{hosseini_hybrid_2019,hosseini_low-mach_2020,hosseini_development_2020}:
\begin{equation}\label{eq:LBM_equation}
 g_i(\bm{r}+\bm{c}_i\delta r,t+\delta t) - g_i(\bm{r},t) = \Omega_i + \delta t\Xi_i,
\end{equation}
where $g_i$ are discrete populations, \change{$\bm{c}_i$ corresponding discrete velocities, $\bm{r}$ and $t$ the position in space and time, $\delta t$ the time-step size }and
\begin{equation}
 \Xi_i = c_s^2\left(f^{\rm eq}_i/\rho -w_i\right)\left(\bm{c}_i-\bm{u}\right)\cdot\bm{\nabla}\rho
 + w_i\rho c_s^2 \Lambda.
\end{equation}
\change{The variable $\Lambda$ appearing in the last term is}
\begin{equation}
 \Lambda = \frac{\partial_t T + \bm{u}\cdot\bm{\nabla}T}{T} \\ + \sum_{k=1}^{N_{\rm sp}}\frac{W}{W_k}\left(\partial_t Y_k + \bm{u}\cdot\bm{\nabla}Y_k\right).
\end{equation}
\change{Here $\bm{u}$ is the fluid velocity vector, $\rho$ the density, $T$ the temperature, $Y_k$ the $k^{\rm th}$ species mass fraction, $W_k$ its molar mass and $W$ the average molar mass, $N_{\rm sp}$ the number of species, $w_i$ the weights associated to each discrete velocity in the lattice Boltzmann solver and $c_s$ the lattice sound speed tied to the time-step and grid size $\delta r$ as $c_s=\delta r/\sqrt{3}\delta t$.} The equilibrium distribution function, $f^{\rm eq}_i$, is given by:
\begin{equation}
 f^{\rm eq}_i = w_i\rho\left(1+\frac{\bm{c}_i.\bm{u}}{c_s^2} + \frac{{(\bm{c}_i.\bm{u})}^2}{2c_s^4} - \frac{\bm{u}^2}{2c_s^2}\right).
\end{equation}
The collision term $\Omega_i$ is defined as:
\begin{equation}
 \Omega_i = -\omega_s\left(g_i-g_i^{\rm eq}\right),
\end{equation}
where 
\begin{equation}
 g_i^{\rm eq} = w_i(P_h - \rho c_s^2) + c_s^2 f_i^{\rm eq},
\end{equation}
and $P_h$ is the hydrodynamic pressure. In the present study first-neighbour stencils based on third-order quadrature are used, i.e. D2Q9 and D3Q27. The \change{hydrodynamic} pressure and momentum are computed as moments of the distribution function $g_i$:
\begin{subequations}
\begin{align}
	P_h &= \sum_{i=1}^{Q} g_i + \frac{\delta t}{2}\rho c_s^2\Lambda,\\
	\rho\bm{u} &= \frac{1}{c_s^2}\sum_{i=1}^{Q} \bm{c}_i g_i.
	\end{align}
\label{eq:moments_PDF}
\end{subequations}

\subsection{Multi-relaxation time realization}
\change{In the context of the present study the Cumulants-based operator is used~\cite{geier_cumulant_2015}. The post-collision populations $g_i^{*}$ are computed as:
\begin{equation}
    g_i^{*} = \rho c_s^2 {f^{'}_i}^{*} + \frac{\delta t}{2}\Xi_i,
\end{equation}
where the post-collision pre-conditioned populations ${f^{'}_i}^{*}$ are:
\begin{equation}
     {f^{'}_i}^{*} = \mathcal{M}^{-1}\left(\mathcal{I} - \mathcal{W}\right)\mathcal{K}^{'} + \mathcal{M}^{-1}\mathcal{W}\mathcal{K}^{'},
\end{equation}
where $\mathcal{M}$ is the moments transform matrix from pre-conditioned populations to the target momentum space, $\mathcal{I}$ the identity matrix and $\mathcal{W}$ the diagonal relaxation frequencies matrix
\begin{equation}
    \mathcal{W}={\rm diag}(\omega_0, \omega_x, \omega_y, ..., \omega_{xxyyzz}),
\end{equation}
where the operator ${\rm diag}$ is defined as:
\begin{equation}
    {\rm diag}(\bm{A}) = (\bm{A}\otimes\bm{1})\circ \mathcal{I},
\end{equation}
with $\bm{A}$ a given vector, $\bm{1}$ a vector with elements 1. The relaxation frequencies of second-order shear moments here shown with $\omega_s$ for the sake of readability, e.g. $xyz$, are defined as:
\begin{equation}
    \omega_s = \frac{\nu}{c_s^2\delta t} + \frac{1}{2}.
\end{equation}
Prior to transformation to momentum space the populations are pre-conditioned as:
\begin{equation}
    f^{'}_i = \frac{1}{\rho c_s^2} g_i + \frac{\delta t}{2\rho c_s^2}\Xi_i.
\end{equation}
This pre-conditioning accomplishes two tasks, namely normalizing the populations with the density and thus eliminating the density-dependence of the moments and introducing the first half of the source term. As such the moments $\mathcal{K}^{'}$ are computed as:
\begin{equation}
    \mathcal{K}^{'}_{j} = \mathcal{M}_{ij} f^{'}_i.
\end{equation}
In the context of the present study we make use of the Cumulants of the distribution function as proposed in~\cite{geier_cumulant_2015}. Details of the collision operator and both forward and backward transforms are given in supplementary materials.}
\subsection{Energy and species mass balance equations}
The species mass and energy balance equations are solved using a finite-difference approximation, with a first-order Euler time-marching scheme. The advection terms are discretized using a third-order weighted essentially non-oscillatory (WENO) scheme while the diffusion terms are treated using a centered second-order approximation. Details on coupling of the different solvers are given in \cite{hosseini_hybrid_2019,hosseini_development_2020}.\\
All thermodynamic and transport parameters are evaluated using the mixture-averaged approximations implemented in the in-house module REGATH. More details on these parameters and the coupling of the solvers to REGATH can be found in~\cite{hosseini_mass-conserving_2018,hosseini_weakly_2020}.
\section{\label{sec:test_case}Numerical application}
\subsection{Deflagrating flame in chamber with obstacles\label{subsec:deflagration}}
The explosion of flammable gases in confined environments is a topic of interest in industrial buildings.
\begin{figure}[!h]
	\centering
		 \includegraphics[width=\columnwidth]{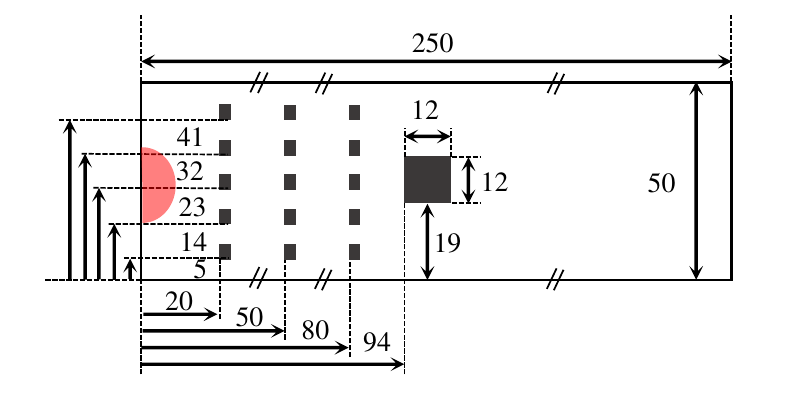}
\caption{Geometrical configuration of deflagration test-case, all dimensions are given in mm.}
\label{Fig:deflagration_2D_geometry}
\end{figure}
Such cases are also rather complicated to model in computational fluid dynamics as they involve a wide spectrum of scales and velocities. To asses the performances of the present model along with boundary conditions we consider the 2-D version of an experimental setup studied in details in \cite{gubba_measurements_2011}, also modeled in \cite{volpiani_large_2017}. 
The configuration consists of a chamber of size $~\hbox{250 mm}\times ~\hbox{50 mm}$. The geometry is illustrated in Fig.~\ref{Fig:deflagration_2D_geometry}. Three rows of five baffle plates, each of size $~\hbox{3 mm}\times ~\hbox{4 mm}$, are placed at a distance of 20~mm from the left wall in the chamber. The horizontal distance between the center of two rows is 30~mm while the vertical distance between baffles is 9~mm . Downstream, at a distance of 94~mm from the left wall there is an obstacle of size $~\hbox{12 mm}\times ~\hbox{12 mm}$. 
\begin{figure}[!h]
	\centering
	\includegraphics[width=0.75\columnwidth]{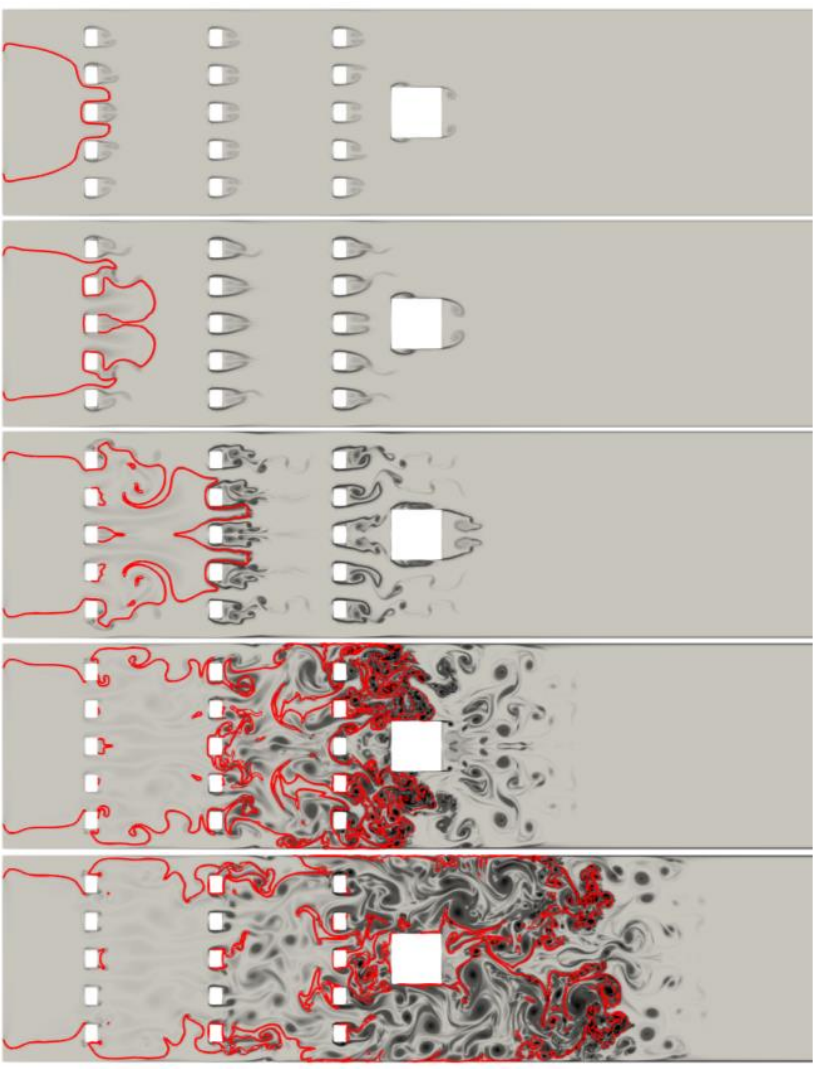}
\caption{Snapshots of the vorticity field (black lines) and flame front (red lines, corresponding to $c=0.2$ and $0.8$) at different times (from top to bottom: 3, 4, 5, 6 and 6.25~ms).}
\label{Fig:chWobstacles_fields}
\end{figure}
Following the simulations in \cite{volpiani_large_2017}, pure propane is used as the fuel. To model propane combustion, we make use of a two-step reduced chemical mechanism proposed in \cite{quillatre_large_2013}. Details on the constants and parameters associated to the two-step scheme can be found in~\cite{quillatre_large_2013,volpiani_large_2017}. The domain is initially filled with a propane/air mixture at equivalence ratio $\phi=1$. A semi-circular area right next to the left wall is filled with burnt gases at the corresponding adiabatic flame temperature (see Fig.~\ref{Fig:deflagration_2D_geometry}). The interface thickness is set to that of the considered mixture. The burnt gas composition, adiabatic temperature and flame thickness are obtained by modeling a 1-D freely propagating flame with the same scheme and mixture. Following~\cite{volpiani_large_2017} the initial radius of the burnt gas area is set to 10~mm. All walls (from side-walls and obstacles) are subject to no-slip/zero-flux/adiabatic boundary conditions.

The LBM simulations are performed on a grid of size $2500\times500$ (i.e., $\delta r=0.1$~mm) so that the flame front is resolved by four points. As the maximum velocity in the domain can grow up to 200~m/s the time-step size is set to $\delta t=5\times10^{-8}$~s, keeping the maximum \change{Courant-Friedrichs-Lewy} (CFL) number in the domain below 0.1. The evolution of the flow field and flame front over time are shown in Fig.~\ref{Fig:chWobstacles_fields}.
\begin{figure}[!h]
	\centering
		\includegraphics[width=0.65\columnwidth]{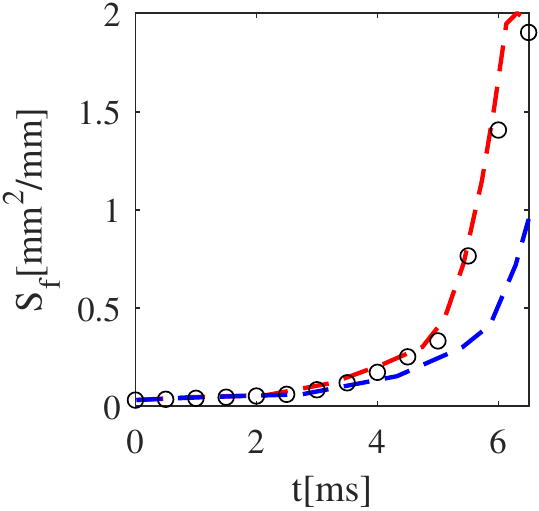}
	\caption{Time-evolution of flame surface in the domain for the 2-D deflagration test-case. DNS and LES data from~\cite{volpiani_large_2017} are shown respectively with red and blue dashed lines while results from the present study are illustrated with black markers.}
	\label{Fig:chWobstacles_surface}
\end{figure}
During the first 3~ms, there is no interaction with the obstacles and the flame front keeps its cylindrical shape. The interaction with the first row of baffle plates accelerates the flame and creates features of relatively large sizes. As the flame progresses through the second and third rows, the flame experiences much larger speeds and smaller vortices. The larger flow speeds and flame surfaces in turn increase the flame propagation speed, which also contributes to a sharp increase of the pressure in the chamber. The evolution of the flame surface in the domain is shown in Fig.~\ref{Fig:chWobstacles_surface}.
There, the results of the present study are compared to DNS and LES data presented in~\cite{volpiani_large_2017}. The grid-size in the former was set to \change{$\delta r=0.07$~mm} while in the latter it was $\delta r=0.14$~mm. The present study is observed to closely follow the former results while the latter exhibits a clear overestimation of the deflagration time.
\subsection{\label{subsec:preccinsta_config}PRECCINSTA burner}
Swirl burners are widely used in many industrial applications as the flow structure induced by the swirler helps stabilize the flame and effectively reduce the size of the system~\cite{benard_large-eddy_2019}.
\begin{figure}[!h]
	\centering
 \includegraphics[width=0.7\columnwidth]{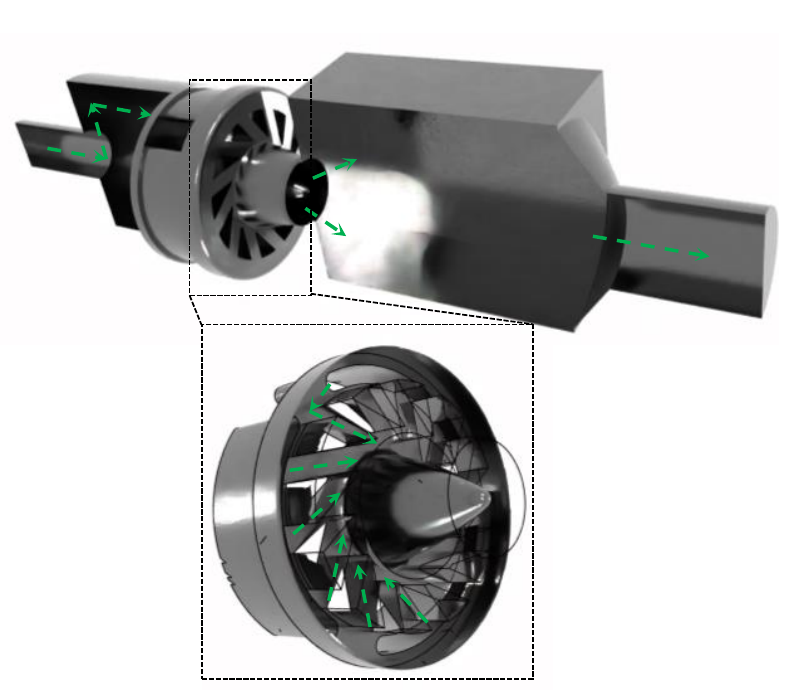}
 \caption{Geometrical configuration of the PRECCINSTA burner.}
\label{Fig:preccinsta_geometry}
\end{figure}
Furthermore, its complex flow structure, geometry, and interaction with the flame front make such burners interesting cases for validation of advanced solvers for combustion simulations. The present section focuses on the numerical study of the well-known PRECCINSTA swirl burner, derived from an industrial design by SAFRAN Helicopter Engines~\cite{benard_large-eddy_2019}.
The geometry involves three main components (Fig.~\ref{Fig:preccinsta_geometry}): (a) a plenum where the mixture is initially injected, (b) the injector made up of twelve radial veins (creating the swirl) and (c) the chamber connected to the injector via a converging nozzle centered around a conical bluff body. The chamber has a size of $86\hbox{ mm}\times86\hbox{ mm}\times110\hbox{ mm}$. The burnt gases get out of the chamber through an exhaust of diameter 39.5~mm~\cite{roux_studies_2005}.
While a wide variety of operating conditions have been considered and studied in the literature~\cite{wang_large_2014,benard_large-eddy_2019,volpiani_large_2017}, the present work follows the set-up presented in \cite{roux_studies_2005}, as it allows for a step-by-step validation of the solver first through a cold flow configuration, before considering the full reacting system.
\subsubsection{\label{subsubsec:results_cold}Results I: cold flow}
\begin{figure*}[h!]
	\centering
 \begin{subfigure}[t]{.32\linewidth}
 \subcaption{Average axial velocity}
 \includegraphics[width=\linewidth]{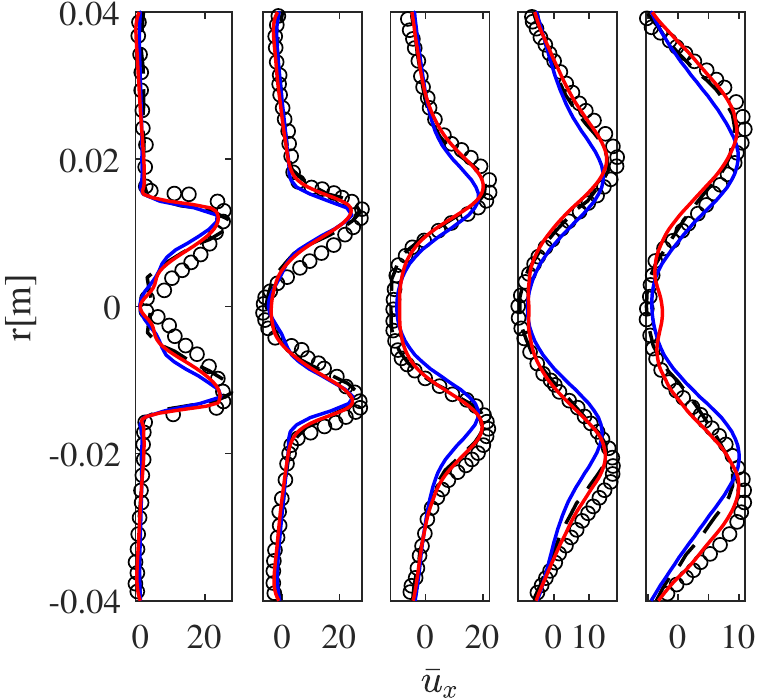}
 \label{Fig:u_avg_axial}
 \end{subfigure}
 \begin{subfigure}[t]{.32\linewidth}
 \subcaption{Average radial velocity}
 \includegraphics[width=\linewidth]{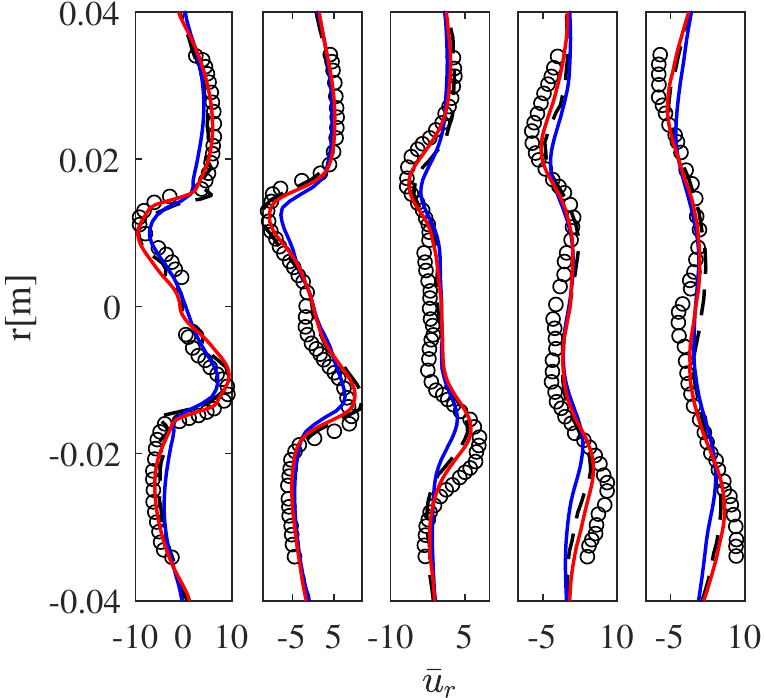}
 \label{Fig:u_avg_radial}
 \end{subfigure}
 \begin{subfigure}[t]{.32\linewidth}
 \subcaption{Average azimuthal velocity}
 \includegraphics[width=\linewidth]{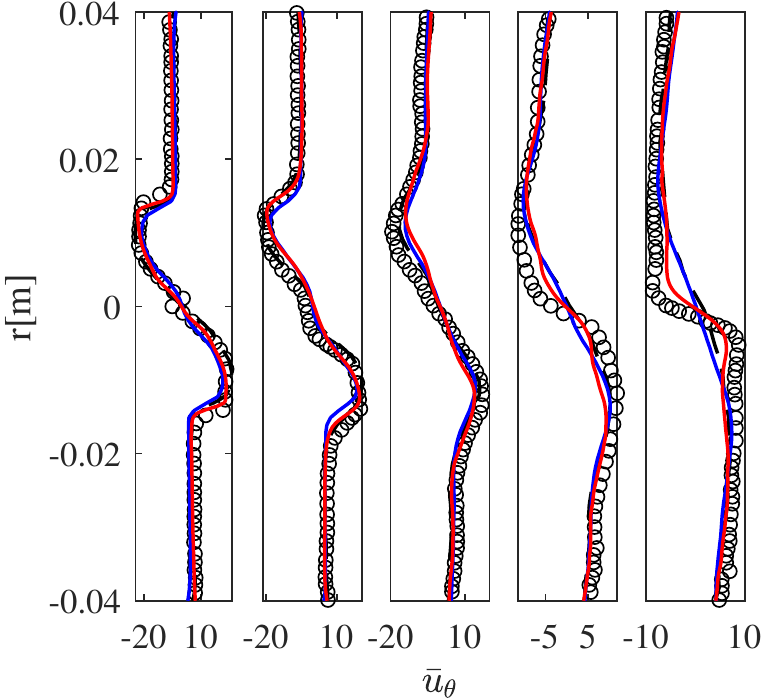}
 \label{Fig:u_avg_azimuthal}
 \end{subfigure}
 \begin{subfigure}[t]{.32\linewidth}
 \subcaption{Root mean square axial velocity}
 \includegraphics[width=\linewidth]{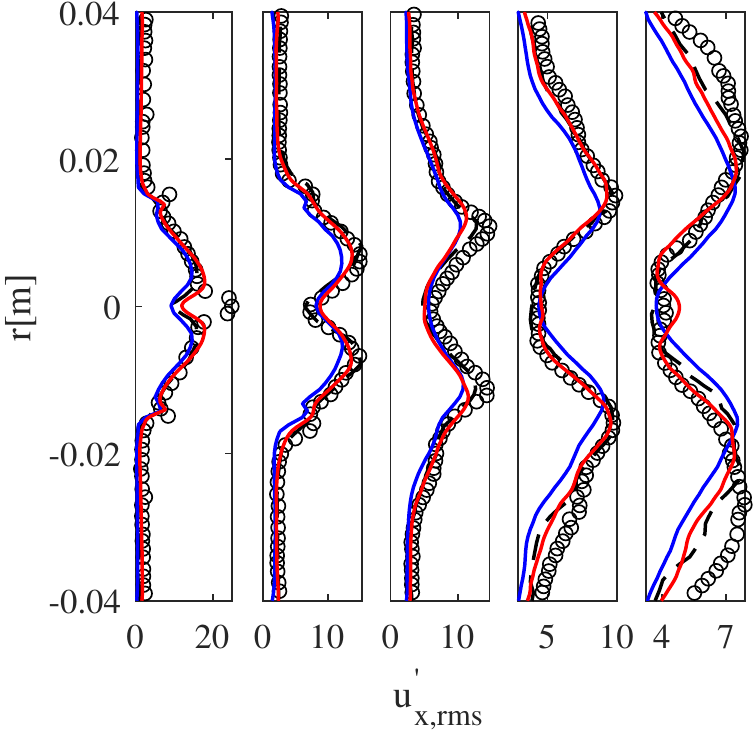}
 \label{Fig:u_rms_axial}
 \end{subfigure}
 \begin{subfigure}[t]{.32\linewidth}
 \subcaption{Root mean square radial velocity}
 \includegraphics[width=\linewidth]{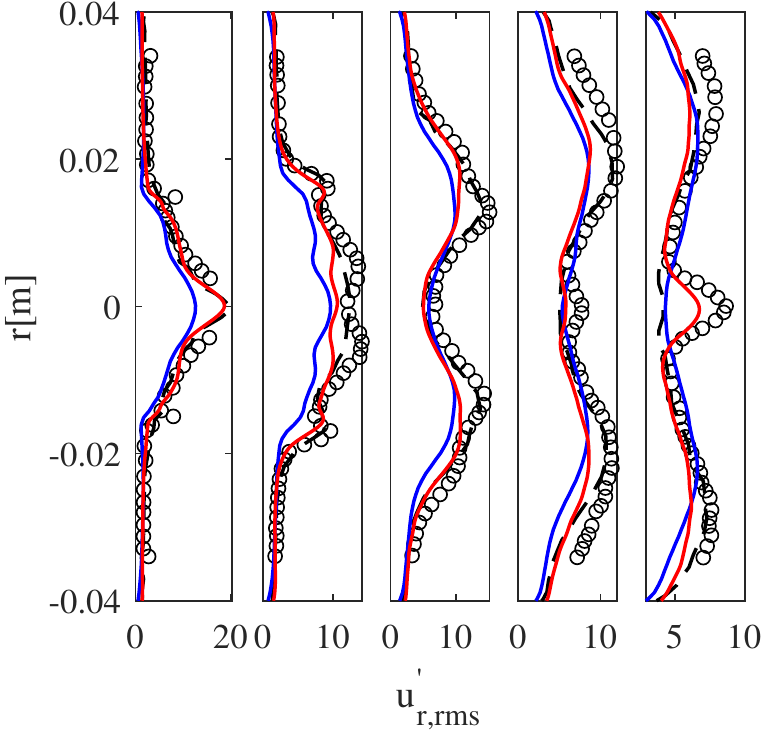}
 \label{Fig:u_rms_radial}
 \end{subfigure}
 \begin{subfigure}[t]{.32\linewidth}
 \subcaption{Root mean square azimuthal velocity}
 \includegraphics[width=\linewidth]{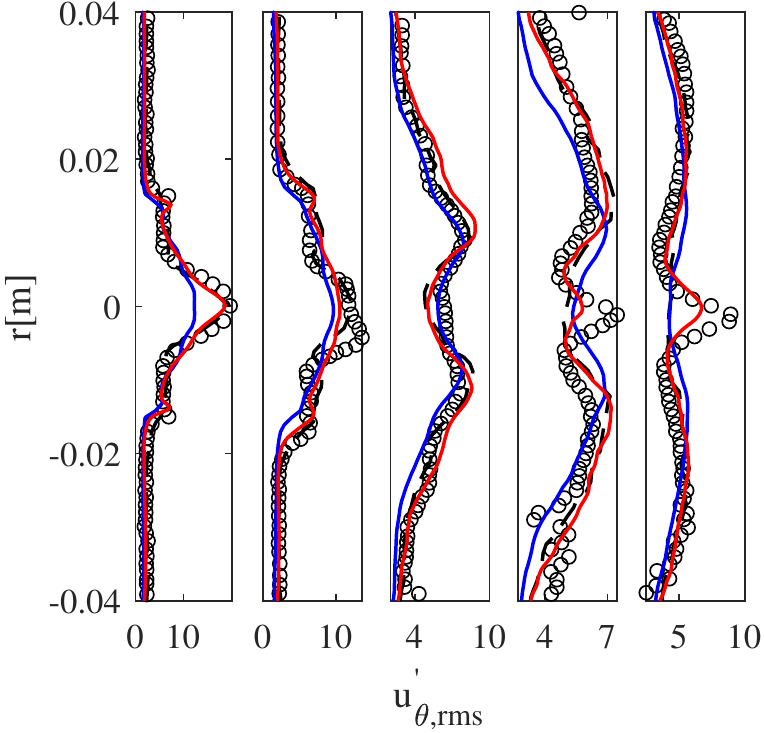}
 \label{Fig:u_rms_azimuthal}
 \end{subfigure}
	\caption{Time-averaged velocity distributions at 5 different distances (from left to right on each panel); 1.5, 5, 15, 25 and 35~mm from the chamber wall. LDA data are shown with with black markers, LES results with dashed black lines while LBM simulations results at resolutions R1 and R2 are respectively represented with blue and red plain lines.}
	\label{Fig:u_cold}
\end{figure*}

\begin{figure}[!h]
	\centering
	\includegraphics[width=\linewidth]{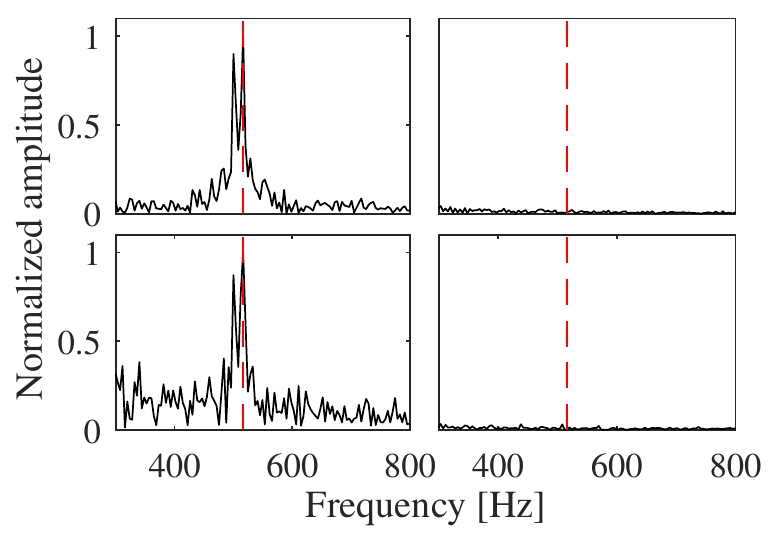}
\caption{\change{(Top) Radial velocity and (bottom) pressure fluctuation spectra for points (left) A and (right) B for the cold flow simulation as obtained from the simulation at resolution R1. The red dashed lines show the PVC frequency, i.e. 516.4 Hz.}}
\label{Fig:PVC_frequency}
\end{figure}

Following~\cite{roux_studies_2005}, the cold flow configuration has a mass flow rate of $12\hbox{ g}.\hbox{s}^{-1}$ at $T=300 \hbox{ K}$ \change{leading to an average inlet velocity of 27 m/s.}
\begin{figure}[!h]
	\centering
	\includegraphics[width=0.75\linewidth]{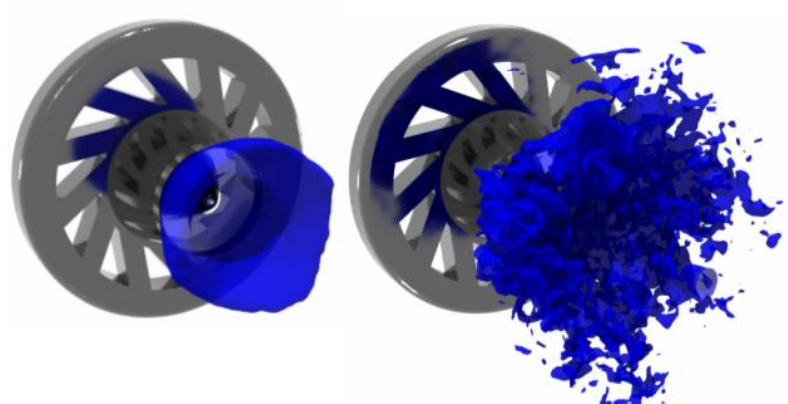}
\caption{Time-averaged (left) and instantaneous (right) velocity fields illustrated via velocity magnitude iso-surfaces (23 m/s)}
\label{Fig:preccinsta_cold_velocity_surfaces}
\end{figure}
The simulations are performed with two different resolutions, (R1) $\delta r=10^{-3}$~m and $\delta t=3\times 10^{-6}$~s, (R2) $\delta r=5\times10^{-4}$~m and $\delta t=1\times 10^{-6}$~s. They are ran for 15 flow-through cycles after which the flow fields are averaged over another 15 cycles, corresponding to approximately $100$~ms of physical time. The velocity fields, both instantaneous and averaged are \change{illustrated} in Fig.~\ref{Fig:preccinsta_cold_velocity_surfaces} via velocity magnitude iso-surfaces.\\
\paragraph{Average velocities and root-mean squares}
The averaged velocity distributions and corresponding root mean squares \change{are first} extracted along five different planes at the following distances from the chamber inlet: 1.5, 5, 15, 25 and 35~mm. The obtained results are compared to both experimental measurements from LDA and numerical data from LES~\cite{roux_studies_2005}. The comparisons are illustrated in Fig.~\ref{Fig:u_cold}.
The plots show that, even though the present LBM simulations are not fully resolved, they exhibit excellent agreement with experimental and LES data.
\paragraph{Root-mean squares of pressure fluctuation in burner}
\change{To further assess the results from LBM simulation the RMS of pressure fluctuations in the burner were extracted at every point. The scatter plot of the pressure fluctuation data is compared to LES data reported in~\cite{roux_studies_2005}.}
\begin{figure}[!h]
	\centering
	\includegraphics[width=\linewidth]{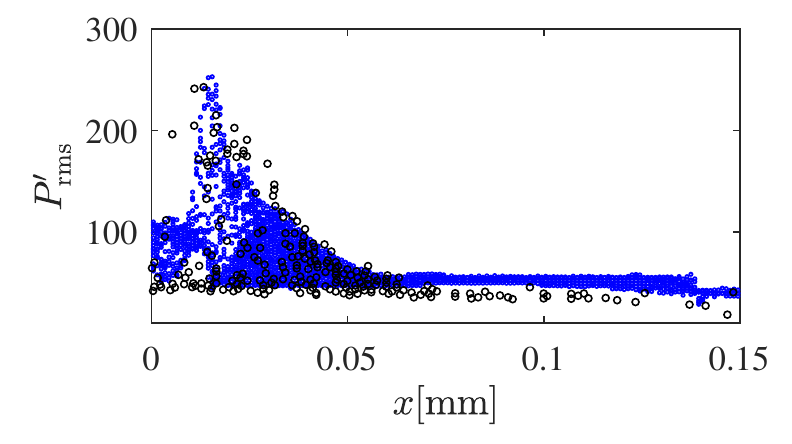}
\caption{\change{RMS of pressure fluctuations (blue dots) as obtained from present simulations and (black markers) extracted from LES reported in~\cite{roux_studies_2005}.}}
\label{Fig:P_rms}
\end{figure}
\change{It is observed that results are in good agreement: The pressure fluctuation peaks close to the swirler inlet at $\approx$250~Pa. The strong fluctuation can be in part attributed to the precessing vortex core formed at the inlet. Further down the burner the fluctuations decrease considerably reaching values between 37 and 50~Pa in the LBM simulations and between 30 and 38 in results reported in~\cite{roux_studies_2005}. }
\paragraph{Precessing vortex core instability} \change{Previous numerical/experimental studies of this configuration have shown that at the inlet of the chamber the swirling flow leads to a precessing vortex core (PVC) instability. The existence of this large-scale hydrodynamic instability is further confirmed by looking at the large RMS fluctuation level around the axis close to the burner inlet. Experimental measurements in~\cite{roux_studies_2005} show an instability at 510~Hz while LES reported in the same publication leads to a hydrodynamic instability at 540~Hz. The PVC is illustrated in Fig.~\ref{Fig:PVC_shape} via isosurfaces of low pressure regions near the burner inlet.
\begin{figure}[!h]
	\centering
	\includegraphics[width=0.6\linewidth]{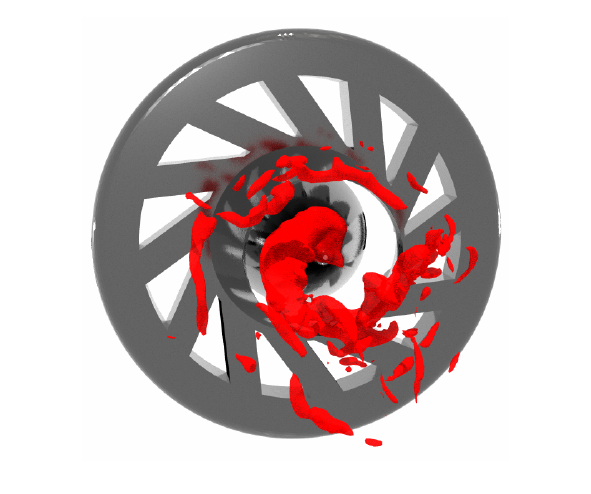}
\caption{\change{Isosurface of low pressure region near the burner inlet as obtained from simulation at resolution R1.}}
\label{Fig:PVC_shape}
\end{figure}
To extract the frequency of the PVC both velocity and pressure were recorded at every time-step over 0.2~s at two different points on the axis at $x_A$=4.1~mm and $x_B$=62.3~mm. The resulting fluctuations along with the corresponding spectra are illustrated in Fig.~\ref{Fig:PVC_frequency}.}
\change{The results as obtained from R1 and R2 simulation are listed in Table~\ref{tab:PVC_frq}.}
\begin{table}[!h]
\centering
\begin{tabular}{||c | c||}
 \hline
    Source & Frequency [Hz] \\
     \hline\hline
    Experiment~\cite{roux_studies_2005} & 510 \\
    \hline
    LES~\cite{roux_studies_2005} & 540 \\
    \hline
    R1 & 516.4 \\
    \hline
    R2 & 524.9 \\
    \hline
\end{tabular}
\caption{\label{tab:PVC_frq}Cold flow PVC frequencies.}
\end{table}
\change{Near the burner inlet, at point A, fluctuations in both pressure and velocity are clearly dominated by the PVC. Further down the burner however, at point B, fluctuations have been considerably damped and the peak at 516.4 has disappeared. This confirms that the peak at 516.4 are limited to the near inlet region and can be attributed to the PVC. It also agrees well with pressure fluctuation levels shown in Fig.~\ref{Fig:P_rms} further confirming that the PVC is the main fluctuation mechanism in chamber.}
\subsubsection{\label{subsubsec:results_hot}Results II: with combustion}

\begin{figure}[!h]
	\centering
	\includegraphics[width=\linewidth]{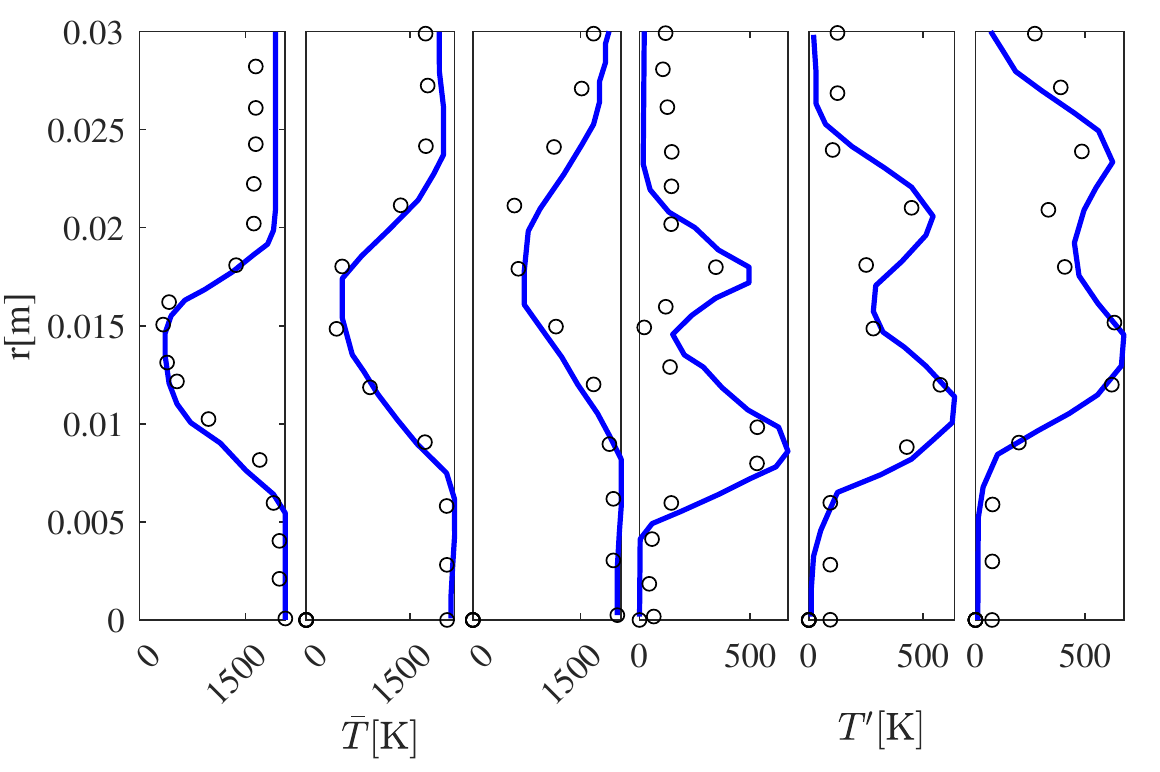}
\caption{\change{Radial distribution of (first three panels) average temperature and (last three panels) RMS of temperature at $x=$6, 10 and 15~mm.}}
\label{Fig:PRECC_hot_T}
\end{figure}

\begin{figure}[!h]
	\centering
	\includegraphics[width=\linewidth]{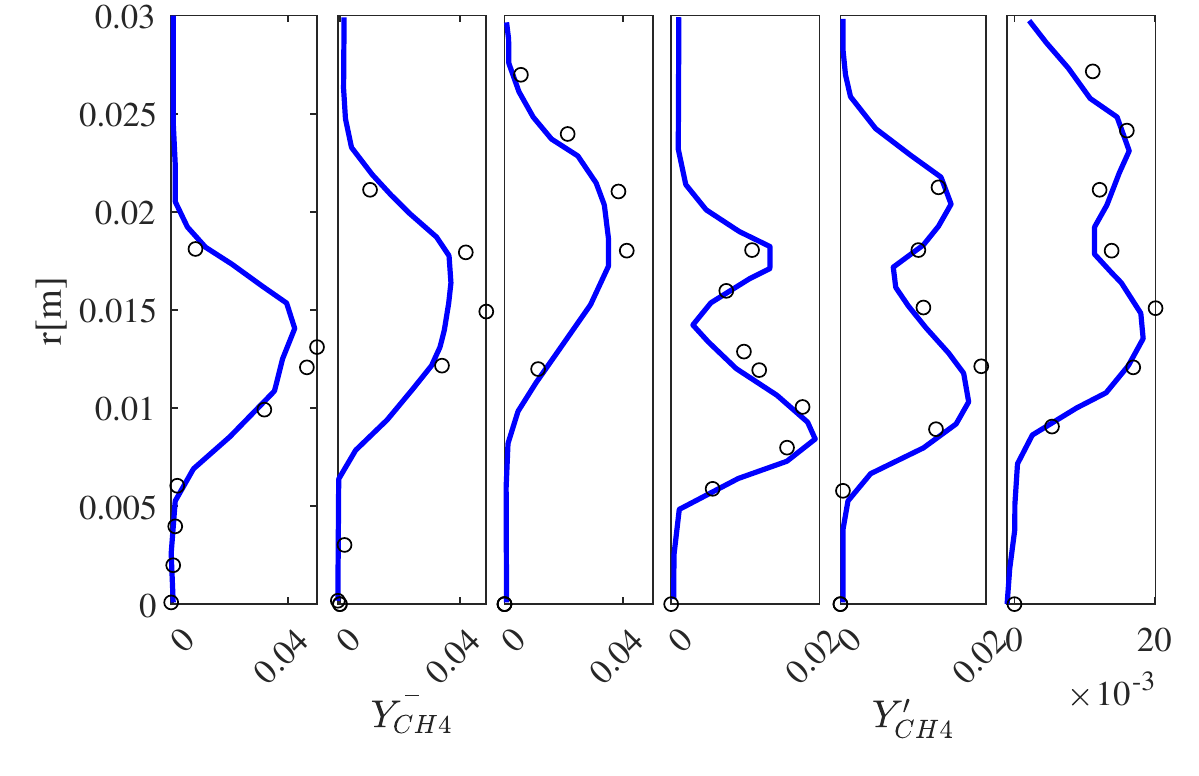}
\caption{\change{Radial distribution of (first three panels) average ${\rm CH}_2$ mass fraction and (last three panels) RMS at $x=$6, 10 and 15~mm.}}
\label{Fig:PRECC_hot_CH4}
\end{figure}

\begin{figure}[!h]
	\centering
	\includegraphics[width=\linewidth]{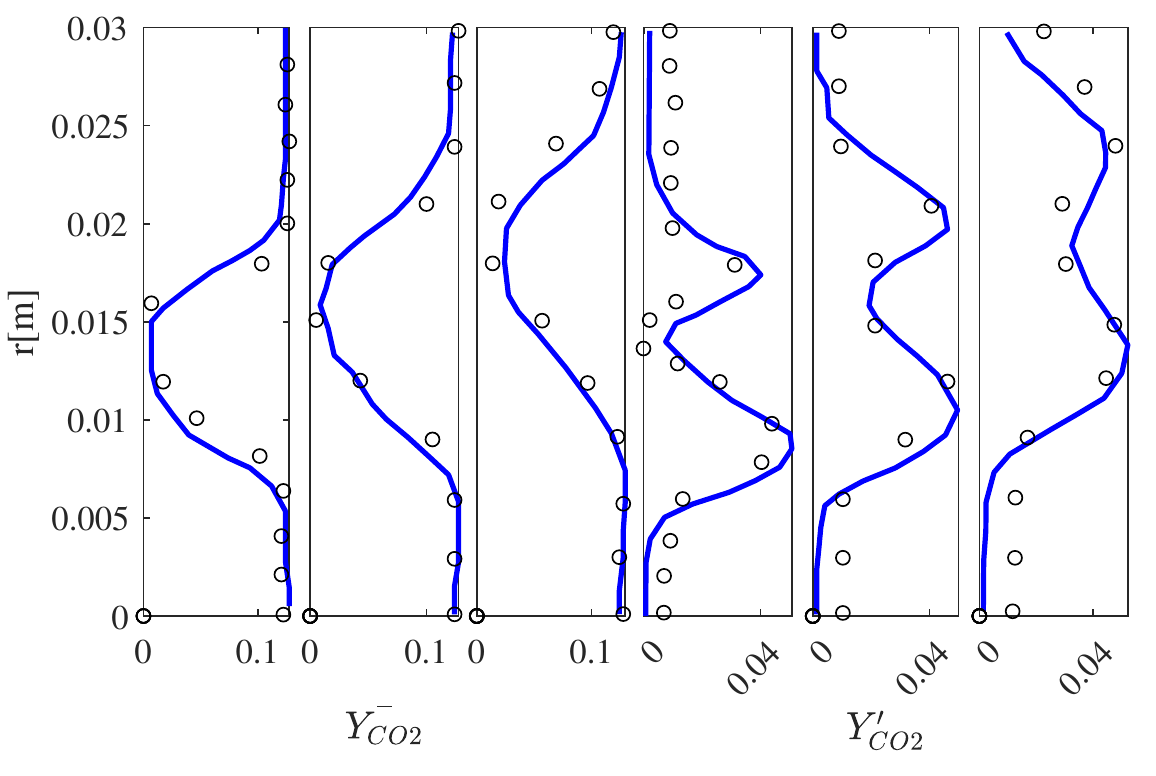}
\caption{\change{Radial distribution of (first three panels) average ${\rm CO}_2$ mass fraction and (last three panels) RMS  at $x=$6, 10 and 15~mm.}}
\label{Fig:PRECC_hot_CO2}
\end{figure}

\change{Following~\cite{roux_studies_2005} we consider here the quiet case of premixed methane/air mixture at equivalence ratio 0.83. Th inlet fresh gas mass flux is $12.9~{\rm g/s}$ at atmospheric pressure and temperature 320~K. This is equivalent to a volumetric flow rate of $0.0122~{\rm m}^3/{\rm s}$ or average velocity of 24.82~m/s. The simulation is conducted using the BFER-2 two-step chemical scheme~\cite{franzelli_impact_2013}. 
For this mixture, the laminar flame speed is found to be $S_L=0.339$~m/s while the flame thickness is computed as 
$
 \delta_L = {T_{\rm max}-T_{\rm min}}/{\max \partial_x T}=0.388~\hbox{mm}.
$
As for the previous configuration the simulation is ran over a total of 30 flow-through cycles, and data is averaged over the second half. The time-step and grid size are set to those used in R2 for the cold configurations and the thickening factor for the flame front is set to $\mathcal{E}=6$. To validate results we compare the mass fraction profiles of $CH_4$ and $CO_2$ and the temperature profile and the corresponding RMS' on three different planes on the $x$-axis at $x=$6, 10 and 15~mm from present simulations and experimental Favre-averaged data extracted from \cite{meier2007detailed}. The obtained data are shown in Figs.~\ref{Fig:PRECC_hot_T}, \ref{Fig:PRECC_hot_CH4} and \ref{Fig:PRECC_hot_CO2}.}
\change{The average temperature and mass fraction profiles along with their corresponding RMS' show good agreement with experimental measurements.}
\section{Conclusions}
\begin{figure}[!h]
	\centering
	\includegraphics[width=0.9\linewidth]{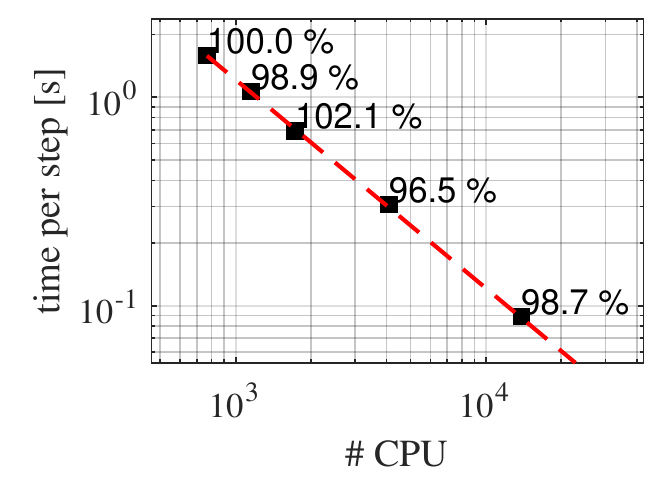}
\caption{\change{Strong scaling data for the PRECCINSTA configuration on superMUC-NG.}}
\label{Fig:Comp_Time}
\end{figure}

A lattice Boltzmann model for low Mach combustion presented in earlier publications was extended for turbulent combustion simulations. The lattice Boltzmann algorithm was supplemented with a Cumulants-based collision operator. The resulting solver was used for the simulation of two complex configurations: 1) deflagration in a 2-D chamber with obstacles, and 2) 3-D PRECCINSTA swirl burner. In both cases, it was shown to correctly model the flow field and flame dynamics. \change{In additional measurements of computation times, as illustrated in Fig.~\ref{Fig:Comp_Time} point to both promising computation times and scaling performances.}
\change{The ALBORZ code performed on average at a rate of 0.4 million lattice updates per second. These performances take account of the both lattice Boltzmann and finite differences solvers and the in-house thermo-chemistry module. It is worth nothing that ALBORZ and REGATH are written in C++ and Fortran respectively. This induced additional communication costs, in the form of multi-array data transformations, pointing to the possibility to further improve performances.\\}
To the authors knowledge this is one of the first and largest (with 30 Mi. grid-points) simulations of turbulent combustion in a realistic burner geometry with a lattice Boltzmann solver for aerodynamic fields. While showing the ability of the numerical model to simulate turbulent combustion, it also opens the way for simulations of combustion in porous media, which is currently under investigation. 
\acknowledgement{Acknowledgments} \addvspace{10pt}
The work was funded (SAH) by the Deutsche Forschungsgemeinschaft (DFG) in TRR 287 (Project-ID 422037413). The authors gratefully acknowledge the Gauss Centre for computation time under grants "pn73ta" and "pn29du" on the GCS Supercomputer SuperMUC-NG at Leibniz Supercomputing Centre, Munich, Germany. SAH thanks S. Volpiani for providing data for the deflagration case \change{(not reported in the literature)} and G. Lartigue and V. Moureau for the PRECCINSTA geometry file.
 \footnotesize
 \baselineskip 9pt
\bibliographystyle{pci}
\bibliography{reference}

\begin{thebibliography}{10}
\expandafter\ifx\csname url\endcsname\relax
  \def\url#1{\texttt{#1}}\fi
\expandafter\ifx\csname urlprefix\endcsname\relax\def\urlprefix{URL }\fi
\expandafter\ifx\csname href\endcsname\relax
  \def\href#1#2{#2} \def\path#1{#1}\fi

\bibitem{yamamoto_simulation_2002}
K.~Yamamoto, X.~He, G.~D. Doolen, Simulation of combustion field with lattice
  {Boltzmann} method, Journal of Statistical Physics 107~(1) (2002) 367--383.

\bibitem{filippova_novel_2000}
O.~Filippova, D.~Hänel, A {Novel} {Lattice} {BGK} {Approach} for {Low} {Mach}
  {Number} {Combustion}, Journal of Computational Physics 158~(2) (2000)
  139--160.

\bibitem{hosseini_hybrid_2019}
S.~A. Hosseini, H.~Safari, N.~Darabiha, D.~Thévenin, M.~Krafczyk, Hybrid
  {Lattice} {Boltzmann}-finite difference model for low mach number combustion
  simulation, Combustion and Flame 209 (2019) 394--404.

\bibitem{tayyab_hybrid_2020}
M.~Tayyab, S.~Zhao, Y.~Feng, P.~Boivin, Hybrid regularized
  {Lattice}-{Boltzmann} modelling of premixed and non-premixed combustion
  processes, Combustion and Flame 211 (2020) 173--184.

\bibitem{sawant_consistent_2021}
N.~Sawant, B.~Dorschner, I.~V. Karlin, Consistent lattice {Boltzmann} model for
  multicomponent mixtures, Journal of Fluid Mechanics 909 (2021).

\bibitem{renard_linear_2021}
F.~Renard, G.~Wissocq, J.-F. Boussuge, P.~Sagaut, A linear stability analysis
  of compressible hybrid lattice {Boltzmann} methods, Journal of Computational
  Physics 446 (2021) 110649.

\bibitem{hosseini_low-mach_2020}
S.~A. Hosseini, A.~Abdelsamie, N.~Darabiha, D.~Thévenin, Low-{Mach} hybrid
  lattice {Boltzmann}-finite difference solver for combustion in complex flows,
  Physics of Fluids 32~(7) (2020) 077105.

\bibitem{tayyab_lattice-boltzmann_2021}
M.~Tayyab, S.~Zhao, P.~Boivin, Lattice-{Boltzmann} modeling of a turbulent
  bluff-body stabilized flame, Physics of Fluids 33~(3) (2021) 031701.

\bibitem{hosseini_development_2020}
S.~A. Hosseini, Development of a lattice {Boltzmann}-based numerical method for
  the simulation of reacting flows, Ph.D. thesis,
  Otto-von-Guericke-Universität/Universit\'e Paris-Saclay (2020).

\bibitem{geier_cumulant_2015}
M.~Geier, M.~Schönherr, A.~Pasquali, M.~Krafczyk, The cumulant lattice
  {Boltzmann} equation in three dimensions: {Theory} and validation, Computers
  \& Mathematics with Applications 70~(4) (2015) 507--547.

\bibitem{hosseini_mass-conserving_2018}
S.~Hosseini, N.~Darabiha, D.~Thévenin, Mass-conserving advection–diffusion
  {Lattice} {Boltzmann} model for multi-species reacting flows, Physica A:
  Statistical Mechanics and its Applications 499 (2018) 40--57.

\bibitem{hosseini_weakly_2020}
S.~Hosseini, A.~Eshghinejadfard, N.~Darabiha, D.~Thévenin, Weakly compressible
  {Lattice} {Boltzmann} simulations of reacting flows with detailed
  thermo-chemical models, Computers \& Mathematics with Applications 79~(1)
  (2020) 141--158.

\bibitem{gubba_measurements_2011}
S.~R. Gubba, S.~S. Ibrahim, W.~Malalasekera, A.~R. Masri, Measurements and
  {LES} calculations of turbulent premixed flame propagation past repeated
  obstacles, Combustion and Flame 158~(12) (2011) 2465--2481.

\bibitem{volpiani_large_2017}
P.~S. Volpiani, T.~Schmitt, D.~Veynante, Large eddy simulation of a turbulent
  swirling premixed flame coupling the {TFLES} model with a dynamic wrinkling
  formulation, Combustion and Flame 180 (2017) 124--135.

\bibitem{quillatre_large_2013}
P.~Quillatre, O.~Vermorel, T.~Poinsot, P.~Ricoux, Large {Eddy} {Simulation} of
  {Vented} {Deflagration}, Industrial \& Engineering Chemistry Research 52~(33)
  (2013) 11414--11423.

\bibitem{benard_large-eddy_2019}
P.~Benard, G.~Lartigue, V.~Moureau, R.~Mercier, Large-{Eddy} {Simulation} of
  the lean-premixed {PRECCINSTA} burner with wall heat loss, Proceedings of the
  Combustion Institute 37~(4) (2019) 5233--5243.

\bibitem{roux_studies_2005}
S.~Roux, G.~Lartigue, T.~Poinsot, U.~Meier, C.~Bérat, Studies of mean and
  unsteady flow in a swirled combustor using experiments, acoustic analysis,
  and large eddy simulations, Combustion and Flame 141~(1-2) (2005) 40--54.

\bibitem{wang_large_2014}
P.~Wang, N.~Platova, J.~Fröhlich, U.~Maas, Large {Eddy} {Simulation} of the
  {PRECCINSTA} burner, International Journal of Heat and Mass Transfer 70
  (2014) 486--495.

\bibitem{franzelli_impact_2013}
B.~Franzelli, E.~Riber, B.~Cuenot, Impact of the chemical description on a
  {Large} {Eddy} {Simulation} of a lean partially premixed swirled flame,
  Comptes Rendus Mécanique 341~(1-2) (2013) 247--256.

\bibitem{meier2007detailed}
W.~Meier, P.~Weigand, X.~R. Duan, R.~Giezendanner-Thoben, Detailed
  characterization of the dynamics of thermoacoustic pulsations in a lean
  premixed swirl flame, Combustion and Flame 150~(1-2) (2007) 2--26.

\end{thebibliography}
\newpage

\small
\baselineskip 10pt

\end{document}